# Parsec-scale Variations in the $^7$Li/$^6$Li Isotope Ratio Toward IC 348 and the Per OB 2 Association[1,1]

D. C. Knauth[2], C. J. Taylor[3,4], A. M. Ritchey[5], S. R. Federman[4], and D. L. Lambert[6]

[1] Based on observations obtained with the Hobby-Eberly Telescope, which is a joint project of the University of Texas at Austin, the Pennsylvania State University, Ludwig-MaximiliansUniversität München, and Georg-August-Universität Göttingen.

[2] 3Woodlawn High School, 1801 Woodlawn Drive, Baltimore, MD 21207; knauth dc2@hotmail.com

[3] Department of Physics and Astronomy, University of Toledo, Toledo, OH 43606; steven.federman@utoledo.edu

[4] Department of Astronomy, University of Maryland, College Park, MD 20742; cjtaylor@astro.umd.edu

[5] Department of Astronomy, University of Washington, Seattle, Washington, 98195; aritchey@astro.washington.edu

[6] 7Department of Astronomy, University of Texas, Austin, TX, 78712; dll@astro.as.utexas.edu

DRAFT: December 4, 2016

[1]Based on data collected at Subaru Telescope, which is operated by the National Astronomical Observatory of Japan.

## ABSTRACT

Measurements of the lithium isotopic ratio in the diffuse interstellar medium from high-resolution spectra of the LiI $\lambda 6708$ resonance doublet have now been reported for a number of lines of sight. The majority of the results for the $^7$Li/$^6$Li ratio are similar to the Solar System ratio of 12.2 but the line of sight toward *o* Per, a star near the star-forming region IC 348, gave a ratio of about 2, the expected value for gas exposed to spallation and fusion reactions driven by cosmic rays. To examine the association of IC 348 with cosmic rays more closely, we measured the lithium isotopic ratio for lines of sight to three stars within a few parsecs of *o* Per. One star, HD 281159, has $^7$Li/$^6$Li ∼ 2 confirming production by cosmic rays. The lithium isotopic ratio toward *o* Per and HD 281159 together with published analyses of the chemistry of interstellar diatomic molecules suggest that the superbubble surrounding IC 348 is the source of the cosmic rays.

*Subject headings:* ISM: atoms — ISM: abundances — ISM: clouds — ISM: structure — Stars: individual (HD 23478, HD 24534 and HD 281159)

## 1. Introduction

In the seminal study of stellar nucleosynthesis by Burbidge et al. (1957), the origin of the two stable lithium isotopes was ascribed to an ill-defined process called the *x*-process. A step toward greater understanding came with the proposal that lithium was a product of Galactic cosmic ray (GCR) spallation and *α*-*α* fusion reactions in interstellar space (Reeves, Fowler & Hoyle 1970; Meneguzzi, Audouze & Reeves 1971, hereafter MAR). Unless the low energy spectrum of cosmic rays is specifically designed to favor $^7$Li production, GCR spallation fails to account for the Solar System value of the $^7$Li/$^6$Li ratio (12.2; Lodders 2003), producing instead a ratio of about two. MAR's alternative to crafting the low energy spectrum was to invoke 'thermonuclear $^7$Li from Giant Stars in "dirty" regions of our Galaxy'.

Today, sources of 'thermonuclear' $^7$Li are known. These sources utilize the CameronFowler mechanism (Cameron & Fowler 1971) in which the reaction $^3$He($\alpha,\gamma$)$^7$Be is followed by $^7$Be(e$^-$,$\nu$)$^7$Li operating in a convective region such that $^3$He is burnt to $^7$Be at 'high' temperatures but the fragile $^7$Be and $^7$Li are convected out to 'low' temperatures and so avoid destruction by protons. Ejection of $^7$Li from stars experiencing the Cameron-Fowler mechanism will raise the interstellar isotopic ratio of lithium from two to a higher value. Known $^7$Li stellar synthesizers include AGB stars (Smith & Lambert 1989), red giant branch stars (Smith et al. 1995; Uttenthaler et al. 2012), novae (Tajitsu et al. 2015; Izzo et al. 2015) and core-collapse supernovae (Woosely & Weaver 1995). In addition, Big Bang nucleosynthesis (BBN) is now recognized as a source of $^7$Li (not $^6$Li), but there remains serious disagreement between the observed and predicted primordial $^7$Li abundance (Fields 2011; Cyburt et al. 2016). This discrepancy influences the need for post-BBN sources of where cosmic rays are prevalent. Observations show that both *o* Per and IC 348 reside within a supernova remnant surrounding the Per OB 2 association (Bally et al. 2008). From measured OH (Snow 1975) and HD column densities (Snow 1976), Hartquist & Morfill (1983) conclude that there is a factor of 10 higher cosmic-ray ionization rate toward *o* Per than toward ζ Per, indicating that the supernova remnant is the source of cosmic rays. These low energy (≤ a few MeV) cosmic rays are produced when ionized material within the supernova remnant is accelerated and focused by the magnetic turbulence in the direction of *o* Per. Based on the radio observations of interstellar OH, Sancisi et al. (1974) find all the OH resides on the near side of *o* Per and conclude that the cosmic rays must originate within the supernova remnant making this star-forming cloud an ideal astrophysical laboratory for further investigations of GCR spallation reactions.

Knauth et al. (2000; 2003) presented moderate and high resolution spectroscopic measurements of the $^7$Li/$^6$Li isotope ratios for the two interstellar clouds observed toward *o* Per in IC 348. The weighted averages of these measurements are $^7$Li/$^6$Li = 1.9 ± 0.3 and 5.9 ± 3.2 in the 4 km s$^{-1}$ and 7 km s$^{-1}$ interstellar clouds, respectively, These low isotope ratios are significantly different than the Solar System value of 12.2 (Lodders 2003) and are consistent with the ratio expected from GCR spallation reactions (MAR).

To determine the extent of these cosmic ray interactions, we acquired data toward additional stars within several parsecs of *o* Per in the active star-forming region IC 348. If the superbubble surrounding IC 348 is the source of the Galactic cosmic rays contributing to the lower isotope ratio toward *o* Per, other nearby sight lines should yield lower than average isotope ratios. Here, we report lithium isotope ratios toward three stars within 15 pc of *o* Per: HD 23478, HD 24534 (X Per) and HD 281159 (BD+31 643).

## 2. Observations and Data Reduction

We obtained high signal-to-noise, high-resolution ($R \geq 100{,}000$) spectra of the Li I doublet at 6708 °A and the K I doublet at 4044 and 4047 °A toward our three target stars. Table 1 shows the relevant stellar data and Figure 1 shows the detections of interstellar Li I and K I toward our target stars.

### 2.1. Subaru

On November 22, 2005, we acquired data toward the three stars using the High Dispersion Spectrograph (HDS) (Noguchi et al. 2002; Sato et. al. 2002) attached to the Subaru Telescope (Iye et al. 2004) located on Mauna Kea, HI. The HDS detector is a mosaic system of two EEV-CCDs, each with 2048 × 4100 pixels, designed to achieve high spectral resolving power and sensitivity, essential for any study of weak interstellar lines. Details of the spectrograph design and performance are provided by Noguchi et al. (2002). For the observations reported herein, we utilized the StdYb setting enabling almost complete wavelength coverage simultaneously from 4140 - 6810 °A with a small gap ∼ 25 °A between the two CCDs. The slit width of the spectrograph was set to 0.2 arcsec, which yielded a spectral resolving power of R ≈ 120,000 as determined from the full width at half maximum (FWHM) of the Th I emission lines in the Th-Ar comparison spectra.

In order to minimize saturation effects and cosmic ray events, we took multiple exposures of HD 23478 and HD 24534 with exposure times ranging from three to thirty minutes. We combined the individual spectra for each star, achieving signal-to-noise ratios (S/N) in the Li I vicinity of 950 and 2680 per resolution element for HD 23478 and HD 24534, respectively. However, for HD 281159 only a single exposure was acquired with a S/N ∼ 100.

## 2.2. HET

We acquired data on HD 281159 with the 9.2 m Hobby-Eberly Telescope (HET) at McDonald Observatory (Shetrone et al. 2007) from August-October 2012. All observations employed the High Resolution Spectrograph (HRS; Tull 1998) with an effective slit width of 125 $\mu$m, which yielded a resolving power of $R\sim100,000$ as determined from the FWHM of Th I emission lines in the Th-Ar comparison spectra. Two spectrographic settings (centered at 4931 Å and 5936 Å) provided data on K I $\lambda$4044 and Li I $\lambda$6708, respectively. For these observations, a total exposure time of 6.6 hr provided a combined S/N $\sim$ 1560 per resolution element.

## 2.3. Data Reduction

Using the NOAO software package, *IRAF*, the acquired spectroscopic data were reduced following a typical prescription for processing ´echelle data. Here we summarize the flow of the data reduction. First, the data were corrected using the median of the 10 bias frames collected during the observing run. Dark noise was determined to be negligible and therefore ignored. Artifacts introduced through cosmic ray interactions and scattered light were removed. The data were flat fielded, with the normalized average of the 20 flat exposures taken throughout the night, to address any variations in the pixel-to-pixel response of the detector. These pixel to pixel variations are minimal as the signal-to-noise scales as the square root of the photon count. Finally, one-dimensional spectra were extracted. Wavelength calibration was performed using identified emission lines in the Th-Ar comparison spectrum. The wavelength calibrated data were shifted to the local standard of rest (LSR) frame and then all data for a given star were co-added together to improve the signal-to-noise ratio. The co-added spectra were trimmed near interstellar species of interest (CH $\lambda$4300, K I $\lambda$4044, and Li I $\lambda$6708) and normalized to unity by fitting low-order polynomials to regions free of interstellar or telluric features. The final co-added spectra for Li I and K I are presented in Figure 1.

## 3. Analysis and Results

For a typical, single, optically thin interstellar cloud that exhibits a Solar System lithium isotope ratio of 12.2, the two fine structure lines of the Li I doublet exhibit relative strengths of 2:1 and a separation of $\sim$ 0.15 ˚A. Coincidentally, the isotope shift and fine structure separation of the lithium atom are such that the stronger $^6$Li absorption line is superimposed on the weaker $^7$Li absorption line. This confusion requires high-resolution and high signal-to-noise spectroscopy (S/N $\sim$1000) to extract meaningful $^7$Li/$^6$Li isotope ratios.

Additional complications arise when multiple clouds are present along the line of sight, which is typical for the interstellar medium (e.g., Welty & Hobbs 2001). Thus, the first step in our analysis was to derive templates for the line-of-sight component structure using the strong $^7$Li I $\lambda$6708, K I $\lambda$4044 and CH $\lambda$4300 lines because the three species are expected to co-exist in an interstellar cloud (Pan et al. 2005). The weaker K I line at 4044 ˚A provides a better template for Li I $\lambda$6708 than CH $\lambda$4300 since both K I and Li I are alkali elements with similar ionization potentials.

Using a C$^{++}$ code, we performed an independent profile synthesis (Knauth et al. 2003) of the Li I, CH, and K I lines and these results are presented in Table 2. In the analysis, the component velocities, $b$-values, and column densities of the species are left as free parameters while minimizing the rms deviations in the residuals of the fit. Hyperfine structure is included in our synthetic profiles for Li I and K I (Knauth et al. 2003), but lambda doubling for CH was not considered for these optically thin lines. The LSR velocity and Doppler broadening parameter for $^7$Li and $^6$Li are assumed to be the same, with the column density of each left as a freely varying parameter. The results from our fits to K I and CH were used to determine the number of velocity components to include in the synthesis for Li I. Our best two component fits to the Li I and K I profiles are presented in Figure 1. The column densities, velocities, and *b*-values resulting from all fits to a given star are in good agreement with each other and the corresponding isotope ratios toward each star are presented in Table 2. Additionally, we performed a consistency check on the Li I profiles using the independently derived CH $\lambda$4300 and K I $\lambda$4044 velocity structure which yield similar $^7$Li/$^6$Li isotope ratios for all three lines of sight in this study.

For HD 23478 and HD 24534, a single component contributes more than 90% of the total column density of Li I, despite the presence of two velocity components detected in both K I and CH at our resolution. However, due to the lower abundance and the inherent uncertainties, only lower limits are presented for the isotope ratio in the second velocity component toward these stars. For HD 281159, the two velocity components observed in both K I and CH are stronger and as a result two components in Li I are clearly evident. In addition, the Li I profile exhibits the fine structure lines of $^7$Li of almost equal depth and not the 2:1 relative strength expected indicating a clear strong presence of $^6$Li along this line of sight. For the stars HD 24534 and HD 281159, the $^7$Li/$^6$Li ratios derived (from K I, CH, and directly fitting Li I) are indistinguishable, however a larger scatter is observed toward HD 23478, which is attributed to the lower S/N of that data, see Table 2. For consistency, we adopt the ratios derived from directly fitting the Li I profile for further analysis.

To gain further insight, we determined the $N$(K I)/$N$(Li I) and elemental K/Li abundance ratios, which are presented in Table 3. We find $N$(K I)/$N$(Li I) = 181 ± 24 toward HD 23478 and slightly higher values of $N$(K I)/$N$(Li I) = 290 ± 56 and $N$(K I)/$N$(Li I) = 408 ± 133 toward HD 24534 and HD 281159, respectively. Following the prescription laid out in Knauth et al. (2003) for HD 24534 and using the extinction curves for HD 23478 (Papaj et al. 1991) and HD 281159 (Snow et al. 1994), we derived the following elemental K/Li ratios: 50 ± 7 toward HD 23478, 81 ± 16 toward HD 24534, and 113 ± 37 toward HD 281159. For comparison, we include our earlier results at the bottom of Table 3 for both $o$ Per (Knauth et al. 2003) and $\zeta$ Per with updated photoionization rates (Knauth et al. 2000; 2003).

## 4. Discussion and Conclusions

The Per OB 2 association has been the subject of numerous investigations due to its close proximity of 316 ± 22 pc (Herbig 1998) and bright stars. These studies reveal that there is a supernova remnant (SNR) surrounding Per OB 2 with $o$ Per inside that shell on the near side of the dense B5 molecular cloud and $\zeta$ Per lying on the far side of the shell (Hartquist & Morfill 1983). The stars $o$ Per and HD 281159 are separated by $8^0$ which corresponds to ~ 0.7 pc at a distance of 300 pc (Luhman et al. 2016). The line of sight to HD 281159 resides near the center of the star-forming region IC 348 (Herbig 1998) with $o$ Per just to the north and HD 23478 about 3 pc beyond IC 348 (Krelowski et al. 1996). $\zeta$ Per and HD 24534 lie more than 10 pc away from $o$ Per.

Analyses of OH and HD observations toward $o$ Per and $\zeta$ Per (Snow 1975, 1976; Hartquist et al. 1978; Federman et al. 1996) reveal an order of magnitude higher cosmic ray ionization rate ($\zeta_p$ = 2.5 × 10$^{-16}$ s$^{-1}$) toward *o* Per than toward $\zeta$ Per ($\zeta_p$ = 2.2 × 10$^{-17}$ s$^{-1}$). Hartquist & Morfill (1983) determine the magnetic field structure of the SNR and claim that the magnetic field focuses the cosmic rays on the dense molecular cloud near *o* Per, thereby enhancing the cosmic ray ionization rate. The ionization rate toward $\zeta$ Per is lower because the magnetic field enters the region at small angles due to the shock enhancing the perpendicular component of the field.

Our results clearly demonstrate that $^6$Li is enhanced within a few parsecs of IC 348. We find a $^1$Li/$^6$Li isotope ratio toward HD 23478 of $^7$Li/$^6$Li = 3.7 ± 0.6 and toward HD 281159, the central star of IC 348, reveal low isotope ratios in both velocity components of $^7$Li/$^6$Li = 1.8 ± 0.8 and 5.1 ± 1.6, which are remarkably similar to the weighted average toward *o* Per (Knauth et al. 2000; 2003) The isotope ratios towards the center of IC 348 (HD 23478, HD 281159, and *o* Per) are lower than the Solar System by factors of 2-4 (see Table 3). While for stars further from IC 348 but still within the Per OB 2 Association, our results for $^7$Li/$^6$Li of 10.6 ± 2.9 toward $\zeta$ Per (Knauth et al. 2000) and of 11.2 ± 4.9 toward HD 24534 are consistent with the Solar System value of 12.2 (Lodders 2003). To date all results are in excellent agreement with the model presented by Hartquist & Morfill (1983).

Indriolo & McCall (2012) measured H$^+_3$ abundances toward stars in this region and determined the cosmic ray ionization rates of $\zeta_2$ = (5.85 ± 3.54) × 10$^{-16}$ s$^{-1}$ toward HD 24534 and (5.55 ± 3.18) × 10$^{-16}$ s$^{-1}$ toward $\zeta$ Per. Above, the primary cosmic ray ionization rate, $\zeta_p$ = 0.5 $\zeta_2$ given in Indriolo & McCall (2012). However, toward *o* Per and HD 281159, Indriolo & McCall (2012) report only upper limits of $\zeta_2$ ≤ 2.55 × 10$^{-16}$ and $\zeta_2$ = ≤ 5.01 × 10$^{-16}$, respectively.

---

[1] Li/$^6$Li ratio toward both *o* Per (Knauth et al. 2000; 2003) and HD 281159 approaches the

Although, these upper limits apparently conflict with our findings, it is important to note that the results of Indriolo & McCall (2012) rely on a number of assumptions concerning the physical conditions in the clouds containing $H_3^+$. More importantly, the abundance of $H_3^+$ traces the current cosmic-ray ionization rate, while the Li isotope ratio probes the integrated cosmic-ray flux. Thus, some differences in these two observables might be expected (Taylor et al. 2012).

If the molecular gas toward $o$ Per and HD 281159 has been subjected to an enhanced flux of cosmic-rays, in accordance with the Hartquist & Morfill (1983) model, then an increase in the relative abundance of $^6$Li would be a natural consequence. Indeed, the value predicted by models of GCR nucleosynthesis (MAR; Ramaty et al. 1997; Lemoine et al. 1998). A decrease in the $^7$Li/$^6$Li ratio due to enhanced cosmic ray activity (resulting in newly-synthesized $^6$Li and $^7$Li), should be accompanied by an increase in the elemental Li abundance. For example, if the clouds toward HD 281159 initially possessed a $^7$Li/$^6$Li ratio of 12.2 (the Solar System value), then the observed ratio of 5.1 implies a 40 % increase in the elemental Li abundance.

Our measurement of the $N$(K I)/$N$(Li I) ratio toward HD 23478 is similar to our earlier results toward $o$ Per and $\zeta$ Per (Knauth et al. 2000; 2003) and to the average value $N$(K I)/$N$(Li I) ' 185 found by Welty & Hobbs (2001) who reported a linear relationship between these two species in diffuse interstellar gas toward 54 galactic stars. While the higher values toward HD 24534 and HD 281159 are unexpected, it may not be appropriate to compare these lines of sight to the diffuse interstellar medium. HD 24534 is an X-ray binary with a circumstellar disk (e.g., Li et al. 2014) and the physical and chemical state of the gas has been altered by processes within IC 348 due to the recent star formation (e.g., Snow et al. 1994). Our determinations of the elemental K/Li abundance toward these lines of sight are similar (see Table 3) to the Solar System average value of 67.6 (Lodders 2003). An alternative interpretation could be that there is either enhanced depletion (or ionization) of Li I and K I in this region.

The discovery of a second low $^7$Li/$^6$Li ratio toward HD 281159, the central star in IC 348, supports the picture that the SNR associated with the Per OB 2 Association is the source of the cosmic ray activity (Knauth et al. 2000; 2003). The observed $^7$Li/$^6$Li ~ 2 as theoretically predicted (MAR) occurs through both alpha-alpha fusion reactions with interstellar He (Fields & Prodanovic 2005) and cosmic rays bombarding C, N, and O nuclei in the interstellar medium surrounding the IC 348 region. Cosmic ray bombardment of the interstellar medium also creates Be and B. It is interesting to note that Ritchey et al. (2011) reported a 50% enhancement in the B II abundance toward *o* Per compared to three other stars in this region (ζ Per, 40 Per, and HD 24534), which provides further support for cosmic ray spallation reactions occurring in this region. However, there remains the unsatisfying result that there is no evidence of enhanced Li I with respect to K I or in the elemental K/Li ratio. Further high-quality observations of the $^7$Li/$^6$Li and $^{11}$B/$^{10}$B isotope ratios toward additional stars in IC 348 and other OB associations (e.g., Sco OB2 and Cyg OB2) are necessary to further quantify light element production via the GCR spallation mechanism.

## 5.  Acknowledgments

We thank Nick Indriolo for useful discussions concerning the cosmic ray ionization rates toward the Perseus stars. Alanna Garay and Akito Tajitsu, the dedicated telescope operators at the Subaru telescope, and the telescope operators at the Hobby-Eberly telescope for their invaluable assistance with collecting these data. DLL thanks the Robert A. Welch Foundation of Houston Texas, through grant F-634. We would like to thank the anonymous referee for their worthwhile comments to our paper.

This paper includes data taken at The McDonald Observatory of The University of Texas at Austin. The Hobby-Eberly Telescope (HET) is a joint project of the University of Texas at Austin, the Pennsylvania State University, Ludwig-Maximilians-Universität München, and Georg-August-Universität Göttingen. The HET is named in honor of its principal benefactors, William P. Hobby Jr. and Robert E. Eberly.

This manuscript was prepared with the AAS LATEX macros v5.2.

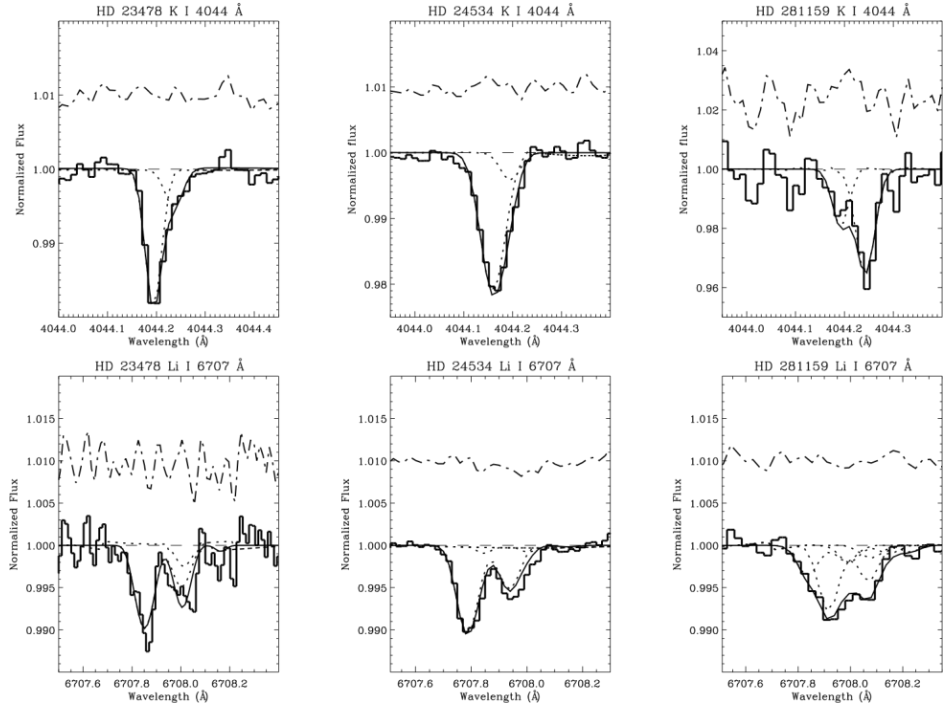

Fig. 1.— The top (bottom) row depicts interstellar K I (Li I) data (solid histogram) toward HD 23478, HD 24534, and HD 281159 normalized to unity. All lines of sight were best fit with two interstellar velocity components seen in K I and Li I (and in CH not presented). The two component fits are shown by the dotted lines. The dot-dashed lines show the residuals to the fit (offset by an appropriate scaling factor), while the dashed lines denote the continuum placement. It is important to note that the $^7$Li I profiles resemble the expected 2:1 intensity ratio towards the stars HD 23478 and HD 24534, but toward HD 281159 the profile is close to 1:1 ratio clearly indicating enhanced $^6$Li along this line of sight.

Table 1.  Stellar Data

| Star | Spec. Type [a] | $V$ [a] (mag) | $B - V$ [a] (mag) | $E(B - V)$ (mag.) | $\tau_{exp}$ (s) | S/N at $\lambda 6708$ | Telescope/Instrument |
|---|---|---|---|---|---|---|---|
| HD 23478 | B3 IV | 6.69 | 0.03 | 0.27 [b] | 7680 | 950 | Subaru/HDS |
| HD 24534 | O9.5 Ve | 6.72 | 0.12 | 0.59 [c] | 6900 | 2680 | Subaru/HDS |
| HD 281159 | B5 | 8.68 | 0.60 | 0.85 [c] | 23690 | 1560 | HET/HRS |

[a]Simbad database, operated at CDS, Strasbourg, France. [b]Papaj, Krelowski, & Wegner 1991. [c]Indriolo & McCall 2012.

Table 2.  Results for Interstellar Species

| Star | Species | Wavelength (Å) | $v_{LSR}$ (km s$^{-1}$) | $b$-value (km s$^{-1}$) | $N$ (cm$^{-2}$) | $^7$Li/$^6$Li [a] |
|---|---|---|---|---|---|---|
| HD 23478 | CH | 4300.313 | 4.0 ± 0.3 | 1.5 ± 0.3 | (1.17 ± 0.18) × 10$^{13}$ | 4.5 ± 1.0 |
|  |  |  | 7.4 ± 0.3 | 1.9 ± 0.3 | (7.08 ± 0.77) × 10$^{12}$ | ≥ 1.1 |
|  | K I | 4044.143 | 3.8 ± 0.1 | 0.4 ± 0.1 | (1.00 ± 0.07) × 10$^{12}$ | 8.3 ± 1.9 |
|  |  |  | 6.9 ± 0.4 | 0.9 ± 0.2 | (2.95 ± 0.44) × 10$^{11}$ | ≥ 4.0 |
|  | $^7$Li | 6707.764 | 3.9 ± 0.3 | 2.2 ± 0.2 | (5.13 ± 0.56) × 10$^9$ | 3.7 ± 0.7 |
|  |  |  | 6.5 ± 0.3 | 3.5 ± 0.3 | (4.37 ± 0.48) × 10$^8$ | ≥ 2.0 |

| Star | Species | Wavelength | | | | |
|---|---|---|---|---|---|---|
| | $^6$Li | 6707.920 | 3.9 ± 0.3 | 2.2 ± 0.2 | (1.38 ± 0.18) × 10$^9$ | ... |
| | | | 6.5 ± 0.3 | 3.5 ± 0.3 | (2.19 ± 0.45) × 10$^8$ | ... |
| HD 24534 | CH | 4300.313 | 1.4 ± 0.3 | 2.3 ± 0.2 | (2.57 ± 0.23) × 10$^{13}$ | 10.7 ± 4.9 |
| | | | 2.9 ± 0.2 | 1.2 ± 0.2 | (9.33 ± 1.01) × 10$^{12}$ | ≥ 3.2 |
| | K I | 4044.143 | 1.2 ± 0.3 | 2.0 ± 0.5 | (1.66 ± 0.15) × 10$^{12}$ | 10.5 ± 4.5 |
| | | | 3.7 ± 0.5 | 1.1 ± 0.1 | (2.57 ± 0.83) × 10$^{11}$ | ≥ 1.1 |
| | $^7$Li | 6707.764 | 1.0 ± 0.3 | 1.8 ± 0.4 | (5.25 ± 0.78) × 10$^9$ | 11.2 ± 5.5 |
| | | | 3.6 ± 0.5 | 0.8 ± 0.1 | (4.47 ± 1.90) × 10$^8$ | ≥ 1.1 |
| | $^6$Li | 6707.920 | 1.0 ± 0.3 | 1.8 ± 0.4 | (4.68 ± 1.99) × 10$^8$ | ... |
| | | | 3.6 ± 0.5 | 0.8 ± 0.1 | (4.37 ± 1.79) × 10$^8$ | ... |
| HD 281159 | CH | 4300.313 | 3.8 ± 0.2 | 1.8 ± 0.2 | (1.32 ± 0.29) × 10$^{13}$ | 1.8 ± 0.8 |
| | | | 7.8 ± 0.2 | 1.7 ± 0.2 | (4.17 ± 0.45) × 10$^{13}$ | 4.9 ± 0.8 |
| | K I | 4044.143 | 3.1 ± 0.6 | 0.7 ± 0.4 | (1.02 ± 0.30) × 10$^{12}$ | 1.7 ± 0.7 |
| | | | 6.9 ± 0.4 | 1.1 ± 0.2 | (2.14 ± 0.36) × 10$^{12}$ | 5.0 ± 0.8 |
| | $^7$Li | 6707.764 | 3.1 ± 0.4 | 0.3 ± 0.4 | (1.86 ± 0.42) × 10$^9$ | 1.8 ± 0.8 |
| | | | 6.8 ± 0.5 | 1.0 ± 0.1 | (4.07 ± 0.36) × 10$^9$ | 5.1 ± 1.6 |
| | $^6$Li | 6707.920 | 3.1 ± 0.4 | 0.3 ± 0.4 | (1.02 ± 0.39) × 10$^9$ | ... |
| | | | 6.8 ± 0.5 | 1.0 ± 0.1 | (7.94 ± 2.32) × 10$^8$ | ... |

[a]Derived using the species, CH, K I, or Li I for profile parameters.

Table 3.    Elemental Abundance Ratios

| Star | $d_{rel}$ [1] | $G(\text{K I})^b$ | $G(\text{Li I})^b$ | $N(\text{K I})/N(\text{Li I})$ | $A(\text{K})/A(\text{L})i$ |
|---|---|---|---|---|---|
| | pc | (s$^{-1}$) | (s$^{-1}$) | | |
| HD 23478 | 2.6 | $1.96 \times 10^{-11}$ | $1.22 \times 10^{-10}$ | 181 ± 24 | 50 ± 7 |
| HD 24534 | 13.9 | $4.89 \times 10^{-12\,c}$ | $3.02 \times 10^{-11\,c}$ | 290 ± 56 | 81 ± 16 |
| HD 281159 | 0.7 | $7.97 \times 10^{-12}$ | $4.98 \times 10^{-11}$ | 408 ± 133 | 113 ± 37 |
| ζ Per $_{c,d}$ | 11.1 | $1.84 \times 10^{-11}$ | $1.15 \times 10^{-10}$ | 191 ± 12 | 53 ± 3 |
| o Per $^c$ | 0.0 | $1.79 \times 10^{-11}$ | $1.11 \times 10^{-10}$ | 186 ± 25 | 52 ± 7 |

[1] Relative distance from *o* Per based on an accepted distance of 300 pc to IC 348 (Luhman et al. 2016).

$^b$Uncertainties on the order of 30%. $^c$Knauth et al. (2003). $^d$Knauth et al. (2000).